\documentclass[twocolumn,preprintnumbers,amsmath,amssymb,showpacs]{revtex4}
\usepackage{dcolumn}
\usepackage{bm}
\usepackage{epsf}

\def\mupair{$\mu^{\pm}$}
\def\eps{\epsilon}
\def\veps{\varepsilon}
\def\pg{$p\gamma~$}

\begin{document}

\title{EeV neutrinos associated with UHECR sources }

\author{Zhuo Li and Eli Waxman}
\affiliation{ Physics Faculty, Weizmann Institute of Science, Rehovot 76100, Israel}

\date{\today}

\begin{abstract}

Electromagnetic energy losses of charged pions and muons suppress the expected high
energy, $\gtrsim10^{18}$~eV, neutrino emission from sources of ultrahigh energy,
$\gtrsim10^{19}$~eV, cosmic-rays. We show here that $\gtrsim10^{19}$~eV photons produced
in such sources by neutral pion decay may escape the sources, thanks to the Klein-Nishina
suppression of the pair production cross section, and produce $\mu^{\pm}$ pairs in
interactions with the cosmic microwave background. The flux of muon decay neutrinos,
which are expected to be associated in time and direction with the electromagnetic
emission from the sources, may reach a few percent of the Waxman-Bahcall bound. Their
detection may allow one to directly identify the sources of $\gtrsim10^{19}$~eV
cosmic-rays, and will provide the most stringent constraints on quantum-gravity-induced
Lorentz violation.

\end{abstract}


\maketitle

\section{Introduction}
The observed characteristics of cosmic rays with energies exceeding $10^{19}$~eV
\cite{nagan000,Bird93,Abbasi05} suggest that there are cosmologically distributed
astrophysical sources that accelerate protons to this high energy~\cite{Waxman_CR_rev}.
The main candidate sources are gamma-ray bursts (GRBs) and active galactic nuclei
(AGN)~\cite{Waxman_CR_rev}. The sources of the ultra-high energy cosmic rays (UHECRs) are
expected to emit also high energy neutrinos, through the decay of charged pions produced
in interactions of high energy protons with the intense radiation fields within the
prospective sources~\cite{w05neutrino}. Large neutrino telescopes, both operating and
under construction, are expected to reach in the coming few years the sensitivity
required for the detection of the predicted high energy neutrino
signal~\cite{neutrino-tel}.

The detection of high energy neutrinos associated
spatially and/or temporally with electromagnetic emission from some sources will provide
evidence for cosmic ray production in these sources. However, electromagnetic and adiabatic
energy losses of charged pions and muons are expected to suppress the neutrino emission at
high energy, where the charged particle life time exceeds its energy loss time~\cite{WB97,Rachen98}.
For example, in the case of GRBs the neutrino flux is expected to be suppressed above a few PeV \cite{WB97,WBbound} to a level which would make it difficult to detect. This would make it
difficult to obtain direct evidence for acceleration of protons
to energy $>10^{19}$~eV.

We discuss here a new type of high energy, $\gtrsim10^{18}$~eV, neutrino signal, which is
expected to be associated with UHECR sources, and which is not suppressed by the energy
losses of charged pions and muons. Neutral pions produced in \pg interactions decay to
photons before undergoing significant adiabatic and/or electromagnetic energy loss, since
their life time is much shorter than that of charged pions and since they are not trapped
by magnetic fields and hardly interact with the plasma and with the radiation field. The
resulting high energy, $\sim10^{19}$~eV, photons are likely to escape the source, due to
the Klein-Nishina suppression of the pair production cross section (see e.g.
\cite{razza04} for a discussion of the GRB case), and may then produce muon pairs in
interactions with cosmic microwave background (CMB) photons. Muon decay leads to the
production of high energy, $\gtrsim10^{18}$~eV, neutrinos, which are associated, both in
direction and in time, with the electromagnetic emission from the sources. High energy
neutrino telescopes like
ARIANNA \cite{arianna} may allow one to detect these $\gtrsim10^{18}$~eV neutrinos.

In what follows we first discuss $\mu^\pm$ pair production in interactions of
$\gtrsim10^{19}$~eV photons with CMB photons, and show that $\sim5$\% of the photon
energy is expected to be converted to (prompt) high energy muons. We then discuss the
production and escape of $\gtrsim10^{19}$~eV photons from GRB sources, as an example of
likely sources of UHECRs which despite being optically thick to pair production for
$1$~GeV to $1$~PeV photons, are likely to be optically thin for $>10^{19}$~eV photons.
Finally, we discuss the expected neutrino flux and spectrum for UHECR sources which, like
GRBs, deposit via pion production a significant fraction of the energy of UHECR protons
in high energy photons, and which are optically thin for $>10^{19}$~eV photons.

\section{$\mu^{\pm}$ pair production in interactions with CMB photons}
Consider a photon of energy $\veps$ propagating through an isotropic radiation field with
photon number density per unit energy ${\rm d}n/{\rm d}\epsilon$. The (inverse) mean free
path (mfp) for $e^\pm$ or $\mu^\pm$ pair production is \cite{WB97}
\begin{equation}\label{eq:mfp_x}
    l^{-1}=\frac{1}{4\varepsilon^2}\int_{2mc^2}^\infty {\rm d}E E^3 \sigma(E)
    \int_{E^2/4\varepsilon}^\infty {\rm d}x x^{-2}
    \frac{{\rm d}n}{{\rm d}\epsilon}(\epsilon=x).
\end{equation}
Here $\sigma(E)$ is the cross section for interaction with energy $E$ in the
center of momentum frame, and $m=m_e$ ($m_\mu$) for $e^\pm$ ($\mu^\pm$) production.
For a thermal radiation of temperature $T$,
${\rm d}n/{\rm d}\epsilon=\pi^{-2}(\hbar c)^{-3}\epsilon^2/[\exp(\epsilon/T)-1]$,
we have, using $u=E^2/4T\varepsilon$,
\begin{equation}\label{eq:mfp_T}
    l^{-1}=\frac{2T^3}{\pi^{2}(\hbar c)^{3}}\int_{(mc^2)^2/T\varepsilon}^\infty {\rm d}u\,
    u \sigma(\sqrt{4 u\varepsilon T})\ln\left[\left(1-e^{-u}\right)^{-1}\right].
\end{equation}
At low energy, $\varepsilon<(mc^2)^2/3T_{\rm CMB}\approx10^{19}$~eV, muon pair production
is possible only in interaction with photons from the high energy tail of the CMB
photon distribution, and the $e^\pm$ pair production mfp, $l_e$, is much shorter
than the $\mu^\pm$ pair production mfp, $l_\mu$. The two mfp's become comparable
at higher energy (note that the cross sections for $\mu^\pm$ and $e^\pm$ production
are similar for  $E>2m_\mu c^2$). Using eq.~(\ref{eq:mfp_T}) we find
$R\equiv l_{\mu}/l_{e}=36,12,8.6$ for $T\veps/(m_\mu c^2)^2=1,3,5$ respectively,
corresponding to $\veps=(0.4,1,2)\times10^{20}(1+z)^{-1}$~eV (Using
$T=T_{\rm CMB}(z)=2.3\times10^{-4}(1+z)$~eV). For photons in the energy range of
$\veps\approx10^{19}(1+z)^{-1}$~eV to $\veps\approx10^{20}(1+z)^{-1}$~eV,
the fraction of high energy photons converted to \mupair~pairs is thus
$f_\mu=1/(1+R)\approx0.03$~to~0.1, and the mfp is $l_{\mu}\approx10^{27}(1+z)^{-3}$~cm.

The fraction of photon energy converted to muons is higher than $f_\mu$, due to the
following reason. The center of momentum energy of the pair production interaction is
well above the electron rest mass. This implies that the $e^\pm$ produced are
relativistic in the center of momentum frame, and that most of the initial photon energy
is carried (in the CMB frame) by one of particles produced. The resulting high energy
electron (or positron) would lose most of its energy in a single inverse-Compton
scattering of a CMB photon (which is deep in the Klein-Nishina regime), leading to the
production of a secondary high energy photon, with energy which is not much smaller than
that of the original photon. This process increases the effective mfp for photon energy
loss to $e^\pm$ to $\simeq 10 l_e$ (e.g. fig.~4 of \cite{sigl}). Since the electron and
photon mfp's are similar, this implies that muon pair production by secondary photons
would increase the fraction of energy loss to muons by a factor of a few beyond $f_\mu$
(The further small additional contribution to muon production via $e\gamma\rightarrow
e\mu^\pm$ may be neglected \cite{MPP2}). As we show below, while the prompt neutrinos
(from the decay of prompt muons produced in an interaction of the original high energy
photons with a CMB photon) are expected to arrive nearly simultaneously with the photons
emitted by the source, neutrinos produced by interactions of secondary photons may suffer
very long time delays due to deflection of the secondary high energy electrons by
inter-Galactic magnetic fields. Since we are interested in neutrinos which may be
associated directly with their sources, we restrict our discussion to neutrinos produced
by the decay of prompt muons.

For the thermal CMB spectrum, $e^\pm$ pair production in interactions with high energy,
$\varepsilon\sim10^{19}$~eV, photons is dominated by photons of energy $\sim T_{\rm
CMB}\sim10^{-3}$~eV, well above the threshold energy, $\sim (m_e
c^2)^2/\varepsilon\sim10^{-7}$~eV. However, the number density of $\sim10^{-7}$~eV
background photons may be dominated by radio emission from objects like galaxies and AGN,
rather than by the CMB. The presence of such additional "radio background" photons may
increase the $e^\pm$ pair production rate, and thus reduce the value of $f_\mu$.
Assuming that the $\sim10^{-7}$~eV ($\sim10$~MHz) radio background observed at Earth is
of extra-Galactic origin, $l_e$ is reduced, with respect to its value in the presence of
CMB photons only, by a factor of $\simeq2$ for $\varepsilon\sim10^{19}$~eV, and by a
factor of $\simeq10$ for $\varepsilon\sim10^{20}$~eV (see, e.g., fig.~1 of \cite{ca97}).
The actual reduction in $l_e$ is expected to be much smaller, since the $\sim10$~MHz
radio background is dominated by Galactic emission. The extra-Galactic contribution to
the radio background is probably lower than 10\% of the observed background (see
discussion in sec.~4 of \cite{Keshet04}), which implies that $l_e$ is not much affected
for $\varepsilon<10^{20}$~eV. Moreover, the contribution of radio background photons to
$e^\pm$ pair production interactions becomes smaller with increasing redshift: At higher
redshift the energy density of the CMB is higher, while the energy density of the radio
background is expected to be lower, since this background is produced by sources like
galaxies and AGN and is therefore accumulating on a time scale comparable to the age of
the universe. Thus, for cosmic-ray sources lying at $z=1$ to $z=2$, which are expected to
dominate neutrino production (e.g. \cite{WBbound}), we expect the effect of radio
background photons to be small.

\section{High energy photon emission: the GRB case}
Both GRBs and AGN may be capable of
accelerating protons to ultra-high energy. The fact that out to $\sim100$~Mpc, the
propagation distance of $10^{20}$~eV protons, there is no AGN luminous enough to meet the
energy output constraint, $L\gtrsim10^{48}{\rm erg/s}$ for proton acceleration to
$10^{20}$~eV \cite{Waxman_CR_rev}, suggests that AGN are not the sources of UHECRs
observed at Earth. We therefore focus below on high energy photon production in GRBs.

Observed GRB properties suggest that they may accelerate protons to $\gtrsim10^{20}$~eV
and that they may produce the UHECR flux observed at Earth \cite{W95prl,waxman95apjl}. In
\cite{WB97} it was shown that $>10^{16}$~eV protons, which are accelerated within the
gamma-ray emitting region of the GRB source, lose a significant fraction, $\sim20$\%, of
their energy to pion production before escaping the source (on average, a proton
undergoes a single pion production interaction before escaping). In fact, for protons
accelerated to $\gtrsim10^{19}$~eV within the gamma-ray emitting region, the fraction of
energy lost to pion production may be somewhat higher than $20\%$. In \cite{WB97}, only
the $\Delta$-resonance contribution to the $p\gamma$ interaction rate was considered.
This is an excellent approximation for $\sim10^{16}$~eV protons, for which the
interaction with the dominant $\epsilon\sim1$~MeV GRB photons is at the
$\Delta$-resonance. Since the spectrum of higher energy photons, for which the
interaction is at an energy higher than the resonance, is steep,
$dn/d\epsilon\propto\epsilon^{-2}$, the contribution of interactions with these photons
to the $p\gamma$ interaction rate is small (see discussion following eq.~1 of
\cite{WB97}). For protons of much higher energy, $\varepsilon\gg10^{16}$~eV, the energy
$\epsilon_\Delta$ of photons for which the interaction is at the $\Delta$-resonance is
lower than 1~MeV, $\epsilon_\Delta\sim 1(10^{16}{\rm eV}/\varepsilon)$~MeV. Since the GRB
spectrum is harder below 1~MeV, $dn/d\epsilon\propto\epsilon^{-1}$, the relative
contribution of non-resonant interactions (compared to the resonant ones) is
$\approx(1/5)\times \ln(\varepsilon/10^{16}{\rm eV})$ (the 1/5 factor is due to the
non-resonant cross section being $\sim5$ times smaller; see eqs.~1 \&~2 of \cite{WB97}).
This holds for protons of energy $\varepsilon<10^{19}$~eV, for which
$\epsilon_\Delta>1$~keV. The GRB spectrum is expected to become self absorbed,
$dn/d\epsilon\propto\epsilon$, at $\epsilon<1$~keV (see below), and $p\gamma$
interactions of $\varepsilon>10^{19}$~eV protons are dominated by the non-resonant
interactions with $>1$~keV photons. The $p\gamma$ interaction rate for these protons is
thus similar to the interaction rate of $\sim10^{16}$~eV protons,
$(1/5)\times\ln(10^{19}/10^{16})\sim1$. However, since the fraction of energy loss per
interaction is higher for interactions well above the resonance, the fraction of energy
loss to pion production may be $\sim50\%$ for $\varepsilon>10^{19}$~eV protons.

Let us consider next the question of whether or not $\gtrsim10^{19}$~eV photons can
escape the GRB source (see also \cite{razza04}). The escape of high energy photons is
limited by pair production interactions. The optical depth to pair production within the
GRB source may be estimated as follows. It is widely accepted that GRB gamma-rays are
produced by the dissipation of the kinetic energy of a highly relativistic outflow,
usually referred to as a "fireball",  driven by accretion onto a compact object
(presumably $\sim1$ solar mass black hole) \cite{GRB_rev}. Let us denote the radius at
which gamma-ray emission arises by $R$, and the outflow Lorentz factor by $\Gamma$. The
pair production optical depth for a photon of high energy $\varepsilon$ is given by the
product of the pair production rate, $1/t'_{\gamma\gamma}(\varepsilon)$, and the
dynamical time, the time required for significant expansion of the plasma, $t'_d\simeq
R/\Gamma c$ (primes denote quantities measured in the outflow rest frame),
$\tau_{\gamma\gamma}(\varepsilon)\simeq R/\Gamma ct'_{\gamma\gamma}(\varepsilon)$.
$t'_{\gamma\gamma}(\varepsilon)$ depends on the energy density and on the spectrum of the
radiation. The (outflow rest frame) radiation energy density is approximately given by
$U'_\gamma=L/4\pi R^2c\Gamma^2$, where $L$ is the GRB luminosity. The GRB spectrum can
typically be described as a broken power law, $dn/d\eps\propto\eps^{-\beta}$, with
$\beta\approx-1$ at low energy, $\eps<\eps_b\sim1$~MeV, and $\beta\approx-2$ at
$\eps>\eps_b\sim1$~MeV. High energy photons with energy $\varepsilon'$ exceeding
$\varepsilon'_b\equiv2(m_ec^2)^2/\eps'_b$, may produce pairs in interactions with photons
of energy exceeding $\eps'=2(m_ec^2)^2/\varepsilon'<\eps'_b$ (the rest frame photon
energy $\varepsilon'$ is related to the observed energy by
$\varepsilon=\Gamma\varepsilon'$). For $\eps'<\eps'_b$ we have
$dn/d\eps'\propto\eps^{\prime-1}$, which implies that the number density of photons with
energy exceeding $\eps'$ depends only weakly on energy. Thus, $t_{\gamma\gamma}$ is
nearly independent of energy for $\varepsilon'>\varepsilon'_b$, $t_{\gamma\gamma}^{\prime
-1}\approx(\sigma_T/16)cU'_\gamma/\eps'_b$, which gives
\begin{equation}\label{eq:tau}
  \tau_{\gamma\gamma}(\varepsilon>\varepsilon_b)\simeq
    10^2\frac{L_{52}}{\Delta t_{-2}\Gamma_{2.5}^4}.
\end{equation}
Here $L=10^{52}L_{52}\rm erg~s^{-1}$, $\Gamma=10^{2.5}\Gamma_{2.5}$,
and the emission radius $R$ is related to the observed
variability time $\Delta t=10^{-2}\Delta t_{-2}$~s by $\Delta t\approx R/2\Gamma^2c$.
Photons of lower energy, $\varepsilon<\varepsilon_b$, interact to produce pairs only
with photons of energy $\eps'>2(m_ec^2)^2/\varepsilon'>\eps'_b$.
Since the number density of these photons drops like $1/\eps'$, we have
$\tau_{\gamma\gamma}(\varepsilon<\varepsilon_b)\approx
(\varepsilon/\varepsilon_b)\tau_{\gamma\gamma}(\varepsilon>\varepsilon_b)$, i.e.
\begin{equation}
  \tau_{\gamma\gamma}(\varepsilon<\varepsilon_b)\simeq
    10^{-3}\frac{L_{52}}{\Delta t_{-2}\Gamma_{2.5}^6}\frac{\varepsilon}{1~\rm MeV}.
\end{equation}
The extension of GRB spectra to $\sim100$~MeV and the characteristic variability time,
$\Delta t_{-2}\sim1$, imply $\Gamma_{2.5}\gtrsim1$. Since thermal pressure acceleration
can not lead to much larger Lorentz factors, $\Gamma_{2.5}\approx1$ is typically adpoted
(e.g. \cite{GRB_rev}).

Eq.~(\ref{eq:tau}) implies that the pair production optical depth is large, $\sim10^2$,
for high energy, $\veps>\veps_b\approx10^{11}\Gamma_{2.5}^2$~eV, photons. However, at
very high energy the source might become optically thin again. This is due to the fact
that the spectra do not extend as $dn/d\eps'\propto\eps^{\prime-1}$ to arbitrarily low
energy. Synchrotron self-absorption is expected to "cutoff" the spectrum at some energy
$\eps'_L\ll\epsilon'_b$, below which the spectrum is expected to follow
$dn/d\eps'\propto\eps^{\prime}$. For high energy photons with energy $\varepsilon'$
exceeding $\veps'_L\equiv2(m_ec^2)^2/\eps'_L$, pair production would be dominated by
interactions with photons of energy $\sim\eps'_L$, and the Klein-Nishina suppression of
the pair production cross section would imply $\tau_{\gamma\gamma}\propto1/\varepsilon'$.
For typical GRB parameters we expect $\epsilon_L\sim1$~keV \cite{LiSong04,LW07},
consistent with BeppoSAX GRB spectra \cite{frontera00}. Using eq.~(\ref{eq:tau}) we thus
have
\begin{equation}
  \tau_{\gamma\gamma}(\varepsilon>\varepsilon_L)\simeq
    10^{-3}\frac{L_{52}}{\Delta t_{-2}\Gamma_{2.5}^2}\left(\frac{\epsilon_L}{1~\rm keV}
    \frac{\varepsilon}{10^{19}\rm eV}\right)^{-1}.
\end{equation}
We note here that the recent detection of (apparently) simultaneous optical/UV and gamma-ray emission
appears to be inconsistent with the standard fireball model, since the optical/UV flux is much
higher than expected in a model where synchrotron self-absorption is important below $\sim1$~keV.
We have shown, however, that the optical/UV emission may in fact be naturally produced in this
model by "residual" dissipation of kinetic energy at radii larger than that of the gamma-ray emission
region \cite{LW07}.

\section{The expected neutrino signal}
Let us assume that the sources of UHECRs deposit a significant fraction of the UHECR
energy output into pion production, and that they are optically thin for $>10^{19}$~eV
photons. The fraction of UHECR energy that would be converted to high energy neutrinos,
via the escape of high energy photons and their interaction with CMB photons, is $\approx
(1/2)\times0.05\times2/3\sim2$\%. The first factor accounts for the fraction of energy
deposited in neutral (as opposed to charged pions), the second accounts for the fraction
of photon energy deposited in (prompt) muon pairs in interaction with CMB photons, and
the last term accounts for the fraction of energy deposited into neutrinos by the muon
decay. This implies that the expected neutrino flux may reach a few percent of the
Waxman-Bahcall bound \cite{WBbound}, which is the neutrino flux that would be obtained if
all the energy produced by the sources of UHECRs were deposited into pions.

At redshift $z$, the energy of photons capable of producing muon pairs in interactions
with CMB photons of energy $\sim 3T_{\rm CMB}$
is $\sim10^{19}(1+z)^{-1}$~eV. The characteristic energy of neutrinos arising from such
interactions is $\sim3\times10^{18}(1+z)^{-1}$~eV, and the energy with which they will be
observed at Earth is $\sim3\times10^{17}[(1+z)/3]^{-2}$~eV.
The neutrino flux from muon production on the CMB is
expected to be strongly suppressed below this energy. The muon decay neutrino flavor
ratio, $\Phi_{\nu_e}^0:\Phi_{\nu_\mu}^0:\Phi_{\nu_\tau}^0=1:1:0$, is expected to be
modified by neutrino oscillations \cite{particle_data} to
$\Phi_{\nu_e}:\Phi_{\nu_\mu}:\Phi_{\nu_\tau}=0.4:0.3:0.3$ at Earth, slightly different
from the $1:1:1$ flavor ratio for neutrinos produced by charged pion decay. The predicted
neutrino flux may be detectable by the suggested ARIANNA neutrino telescopes
\cite{arianna}, which is designed for detection of neutrinos of energy
$\gtrsim10^{17.5}$~eV.

It is important to emphasize here that the (prompt) neutrino signal is expected to be
associated in both time and direction with the electromagnetic emission of the source,
and that this association can not be affected by the presence of inter-galactic magnetic
fields, which may deflect the charged muons. The amplitude of a large scale
inter-galactic magnetic field is limited by Faraday rotation measures to less than 1~nG
\cite{IGMF}. Given the muon rest-frame lifetime, $t_\mu=2\times10^{-6}$~s, the angle by
which the muon may be deflected before decaying is only $\theta\approx t_\mu eB/m_\mu
c\sim10^{-10}(B/1{\rm nG})$, and the related time delay is $\Delta
t_B\approx\theta^2l_{\mu}/c\sim1(B/1{\rm nG})(l_{\mu}/10^{27}{\rm cm})$~ms. On the other
hand, neutrinos produced by the decay of secondary muons (muons produced by high energy
photons generated by inverse-Compton scattering of CMB photons by $e^\pm$ pairs produced
by the original high energy photon), may suffer strong delays since the mean free path of
secondary electrons is large, $\sim10^{26}$~cm, and they may therefore be significantly
deflected by inter-Galactic fields.

The comparison of arrival times of GRB-induced neutrinos and GRB photons may allow one to
test for Lorentz invariance violation (LIV) as expected due to quantum gravity effects
\cite{WB97,LIV}. Since the time delay is expected to be an increasing function of
neutrino energy, $\Delta t_{\rm LIV}\approx(d/c)(\eps_\nu/\xi E_{\rm planck})^n$, the
$10^{18}$~eV neutrinos discussed in this paper would provide much more stringent tests
than the pion decay $\sim10^{15}$~eV neutrinos expected to be produced within the GRB
source. The accuracy with which the time delay could be measured is limited by the
duration of the burst photons, which ranges from few ms's for short bursts to tens of
seconds for long ones. A possible complication in this context is the possible emission
of $\sim10^{17}$~eV neutrinos during the GRB "afterglow" phase, which may take place on
time scale of hours \cite{Li02}. For $10^{18}$~eV neutrinos originating from $z=1$
($d\sim10^{28}$~cm), a time delay of hours would probe $\xi_1\gtrsim10^4$ for $n=1$ or
$\xi_2\gtrsim10^{-3}$ for $n=2$.  It is important to note here that since the background
level of neutrino telescopes is very low, especially at high neutrino energy, the
detection of a single EeV neutrino from the direction of a GRB on a time sale of even
years after the burst would imply an association of the neutrino with the burst and will
therefore establish a time delay measurement.

ZL thanks Boaz Katz for helpful discussion. This work was partially supported by Minerva and ISF
grants.


\begin{thebibliography}{99}

\bibitem{nagan000} M. Nagano and A.A. Watson, Rev. Mod. Phys. {\bf 72}, 689 (2000)

\bibitem[Bird et al.(1993)]{Bird93}
  D.~Bird, et. al.  \prl {\bf71}, 3401 (1993).

\bibitem[Abbasi et al.(2005)]{Abbasi05}
  R.~U.~Abbasi, et al.\  \apj, {\bf622}, 910 (2005).

\bibitem{Waxman_CR_rev}
  E.~Waxman, Pramana {\bf62}, 483 (2004) [arXiv:astro-ph/0310079].

\bibitem[Waxman(2005)]{w05neutrino} E. Waxman,\  Physica
Scripta Volume T, {\bf 121}, 147 (2005)

\bibitem{neutrino-tel} F. Halzen, Proc. Nobel Symp. 129: Neutrino Physics (World Scientific, Enkping, 2004),
[arXiv:astro-ph/0501593].

\bibitem[Rachen \& M{\'e}sz{\'a}ros(1998)]{Rachen98} J. P. Rachen,
and P. M{\'e}sz{\'a}ros,  \prd {\bf 58}, 123005 (1998)

\bibitem{WB97}E. Waxman and J. N. Bahcall, Phys. Rev. Lett. {\bf 78}, 2292 (1997).

\bibitem[Waxman \& Bahcall(1999)]{WBbound} E. Waxman and J. N. Bahcall, \prd {\bf 59}, 023002 (1999)

\bibitem[Razzaque et al.(2004)]{razza04}  S. Razzaque, P.,
M{\'e}sz{\'a}ros, and B. Zhang, \apj, {\bf 613}, 1072 (2004)

\bibitem[Barwick(2006)]{arianna} S.~W. Barwick,\ 2006, arXiv:astro-ph/0610631

\bibitem[Sigl(2001)]{sigl} G. Sigl,\  Physics and
Astrophysics of Ultra-High-Energy Cosmic Rays, {\bf 576}, 196 (2001)

\bibitem[Athar et al.(2001)]{MPP2} H. Athar, G.-L. Lin, and J.-J. Tseng, \prd {\bf
64}, 071302 (2001).

\bibitem[Coppi \& Aharonian(1997)]{ca97}  P.~S. Coppi and F.~A.
Aharonian, \apj {\bf 487}, L9 (1997)

\bibitem[Keshet et al.(2004)]{Keshet04} U. Keshet, E. Waxman,
and A. Loeb, \apj {\bf 617}, 281 (2004)

\bibitem{W95prl}E. Waxman, Phys. Rev. Lett. {\bf 75}, 386 (1995).

\bibitem[Waxman(1995)]{waxman95apjl} E. Waxman  \apj {\bf 452},
L1 (1995).

\bibitem{GRB_rev} B. Zhang and P. M{\'e}sz{\'a}ros, Int. J. Mod. Phys. A {\bf 19}, 2385
(2004); B. Zhang, Chinese J. Astron. Astrophys. {\bf 7}, 1 (2007)

\bibitem[Li \& Song(2004)]{LiSong04} Z. Li and L. M. Song,
\apj {\bf 608}, L17 (2004)

\bibitem[Li \& Waxman(2007)]{LW07} Z. Li and E. Waxman,
2007, arXiv:0711.2379

\bibitem[Frontera et al.(2000)]{frontera00} F. Frontera, et al.\
 \apj Supp.  {\bf 127}, 59 (2000)

\bibitem[Particle Data Group et al.(2004)]{particle_data} Particle
Data Group,  Physics Letters B, {\bf 592}, 1 (2004)


\bibitem[Kronberg(1994)]{IGMF} P.~P. Kronberg,
Reports of Progress in Physics, {\bf 57}, 325 (1994)

\bibitem{LIV} G. Amelino-Camelia, et al. \nat {\bf 393}, 763 (1998); S. Coleman and S. L. Glashow,
\prd {\bf 59}, 116008 (1999); U. Jacob and T. Piran, Nature Phys. {\bf 3}, 87 (2007).

\bibitem{Li02} Z. Li, Z. G. Dai and T. Lu, Astron. Astrophys. {\bf 396}, 303 (2002);
C. D. Dermer, Astrophys. J. {\bf 574}, 65 (2002).

\end{thebibliography}
\end{document}